# *Electrical and Structural Response of Nine-Atom-Wide Armchair Graphene Nanoribbon Transistors to Gamma Irradiation*

*Kentaro Yumigeta[a,#], Muhammed Yusufoglu[a,#], John G. Federice[b], Elena T. Hughes[b], Ahmet Mert Degirmenci[c], Jon T. Njardarson[b], Kelly Simmons-Potter[a,c,d], Barrett G. Potter[a,d,\*], and Zafer Mutlu[a,c,e,\**

[a] *Department of Materials Science & Engineering, University of Arizona, Tucson, Arizona 85721, USA*

[b] *Department of Chemistry and Biochemistry, University of Arizona, Tucson, Arizona 85721, USA*

[c] *Department of Electrical and Computer Engineering, University of Arizona, Tucson, Arizona 85721, USA*

[d] *James C. Wyant College of Optical Sciences, University of Arizona, Tucson, Arizona 85721,*

[e] *Department of Physics, University of Arizona, Tucson, Arizona 85721, USA*

[#] *Kentaro Yumigeta and Muhammed Yusufoglu contributed equally to this study.*

\* *Corresponding Authors: zmutlu@arizona.edu, bgpotter@arizona.edu*

## Abstract

Materials and devices used in space and advanced energy systems are continuously exposed to high-energy photons and particles, leading to gradual changes in their structural and electronic properties. Gamma-ray exposure is particularly critical because their strong penetrating power allows them to traverse conventional shielding and device packaging. Real-time monitoring of exposure-induced changes in compact, chip-integrated devices remains limited despite the availability of external radiation detectors. Atomically precise graphene nanoribbons (GNRs) present an attractive platform for probing such effects due to their structural uniformity, tunable electronic properties, and exceptional sensitivity of charge transport to even subtle lattice modifications, a capability not yet demonstrated in other low-dimensional materials. Here, we investigate the structural and electronic response of atomically precise GNRs under gamma irradiation. Nine-atom-wide armchair GNRs (9-AGNRs) were synthesized via a bottom-up on-surface approach, integrated into field-effect transistors (FETs), and characterized before and after exposure using Raman spectroscopy and electrical transport measurements. Raman spectroscopy indicates preservation of the primary GNR lattice structure, accompanied by subtle spectral changes suggestive of irradiation-induced oxidation or local lattice perturbations. While these measurements do not indicate severe structural damage, electrical transport measurements reveal a pronounced degradation in device performance, demonstrating the strong susceptibility of GNRFETs to gamma-ray exposure. This pronounced response may be attributed to Anderson localization of charge carriers, potentially arising from enhanced quantum interference in atomically narrow, quasi-one-dimensional GNRs. These results highlight the potential of GNR-based nanoelectronic devices for sensing and monitoring under extreme operational conditions.

## Introduction

The advancement of electronic and structural systems for demanding environments, such as deep-space exploration, high-performance aerospace, and next-generation energy technologies, fundamentally relies on materials and devices capable of continuously monitoring their structural and electronic state-of-health. Reliable assessment of system integrity in these settings is critical for guiding autonomous operational decisions and enabling timely adaptive responses to environmental changes.

In space, for instance, electronic components are exposed to energetic particle fluxes and high-energy photons that can modify material properties and affect device performance over time.[1,2] Although in-situ monitoring of these radiation-induced changes is critical for assessing device reliability, conventional microelectronic platforms are often not sufficiently sensitive or robust to detect the subtle changes induced by extreme environments and compact form factors required for chip-integrated monitoring.[3,4] This critical gap motivates the development of new materials and device architectures that can effectively convert environmental stress into a measurable electrical signal.

Among emerging material platforms, low-dimensional nanomaterials, including 2D materials, carbon nanotubes (CNTs), and graphene, are particularly suited for this purpose, offering unique advantages for nanoscale sensing.[5,6] Their high surface-to-volume ratio, combined with exceptional mechanical and thermal stability, makes them promising building blocks for reliable nanoscale devices.[7,8] Quantum confinement in these materials also produces electronic properties that differ significantly from bulk counterparts, allowing desired electronic characteristics to be engineered through rational design of the material structure.[9]

However, this strong dependence of electronic properties on atomic-scale structure means that achieving reproducible device performance requires precise control during fabrication to minimize inhomogeneities. Atomically precise graphene nanoribbons (GNRs) address need for structural uniformity through a bottom-up synthesis approach wherein a single organic monomer undergoes homopolymerization to form GNRs with uniform, well-defined width and edge structure.[10–12] This synthetic molecular design strategy ensures that all resulting GNRs are structurally identical, providing a uniform baseline against which their quantum-confined charge transport remains highly sensitive to even small perturbations in GNR structure brought about by external forces and chemical reactions. As a result, modifications at the edges or within the lattice produce clear, measurable changes in electronic behavior. To date, atomically precise synthesis that produces low-dimensional conductors with a uniform atomic structure encoded by monomer design, including width and edge termination, has not been achieved for other low-dimensional material platforms such as conventional 2D semiconductors or carbon nanotubes. This structural uniformity and exceptional sensitivity to small changes in GNR structures make them well-suited for monitoring both device state and nanoscale environmental conditions.[13]

Despite this promising potential, the effects of exposure to high-energy irradiation on the structural and electronic properties of atomically precise GNRs remain unexplored. Gamma-ray exposure is a frequent cause of radiation-induced degradation in electronic systems due to its high penetration depth.[1–4]

Understanding these fundamental interactions between gamma rays and GNRs is essential for assessing GNRs potential as reliable, sensitive nanoscale sensors in extreme environments. Establishing the link between any GNR structural changes brought about upon gamma irradiation and the resulting impact on electronic behavior provides a necessary foundation for evaluating their suitability in high-impact applications.

In this study, we investigate defect formation induced by gamma irradiation in GNRs and its impact on electrical characteristics of nine-atom-wide armchair graphene nanoribbon (9-AGNR) field-effect transistors (FETs). To monitor the impact of gamma irradiation on these devices, we employed Raman spectroscopy to provide insights into changes in molecular structure, as well as electrical transport measurements to assess variations in device performance, both before and after irradiation. We present possible origins of the observed structural and electronic changes. This work represents an initial step toward establishing the structural and electronic changes of exposure to high-energy irradiation on atomically precise graphene nanoribbons, providing insight into their relative resilience and helping to guide future studies of radiation–matter interactions and potential device-relevant applications.

## Results and Discussion

9-AGNRs were synthesized on Au(111)/mica substrates via on-surface synthesis by homopolymerization of 3′,6′-diiodo-1,1′:2′,1″-terphenyl (DITP),[14] and aromatization upon further heating route (**Figure 1a**), with detailed synthesis of DITP provided in the Supporting Information. Specifically, DITP monomer was deposited onto the Au substrate at room temperature, followed by annealing at 200 °C to induce homopolymerization. The temperature was subsequently raised to 400 °C to promote cyclodehydrogenation, forming GNRs.

Raman spectroscopy (**Figure 1b**) was used to assess the structural integrity of the as-grown GNRs, as it is well suited for probing characteristic vibrational modes in carbon materials and evaluating their degree of structural order.[15–19] The 9-AGNR samples exhibited sharp, well-defined peaks characteristic of planar GNRs, including a set of features in the G-band region (near ~1600 cm$^{-1}$) associated with sp$^2$ C–C stretching vibrations. For simplicity, we refer to the dominant peak in this region as the "G peak". The spectrum clearly features the radial breathing-like mode (RBLM) in the low-frequency region (~396 cm$^{-1}$). Because the RBLM frequency is highly sensitive to ribbon width,[10,12,20,21] the observed RBLM peak provides supporting evidence for the targeted 9-AGNR structure. Peak deconvolution and fitting details for the Raman spectra are provided in the Supporting Information.

Regarding other prominent Raman features, the peak near 1340–1350 cm$^{-1}$ is widely known in the 2D graphene field as the defect-activated D peak.[15–19] This peak arises from a double-resonance process that requires defect scattering to satisfy momentum conservation. In GNRs, the finite size and the intrinsic edges break translational symmetry and can provide the necessary scattering. Therefore, the D band is not necessarily indicative of additional point defects and may be observed even for structurally well-defined GNRs. In addition, peaks around ~1250 cm$^{-1}$ have conventionally been assigned to C–H in-plane bending modes associated with hydrogen-terminated ribbon edges (i.e., CH modes).[18,19]

However, recent studies have provided a revised interpretation for the 1200–1400 cm$^{-1}$ region in atomically precise GNRs.[22] These bands originate from phonon branches that are Raman-inactive in 2D graphene but become Raman-active in GNRs due to lateral quantum confinement. As detailed in the Supporting Information using 2D graphene phonon dispersion, the nanometer-scale width of 9-AGNR imposes quantization of the phonon wavevector in the ribbon width direction, which folds phonon branches and activates otherwise-forbidden modes. In this work, we use the abbreviation "CH/D" to collectively denote the features in the D-band and C–H-related spectral regions.

Although the ~1335 cm$^{-1}$ peak can be intrinsically present in pristine GNRs due to confinement-related activation, its intensity is still enhanced by additional structural disorder. Therefore, the intensity ratio between the ~1335 cm$^{-1}$ peak and the G peak at ~1600 cm$^{-1}$ (denoted as $I_D/I_G$ following convention) remains a valid quantitative metric for evaluating the defect level.[23–25] In our samples, the measured $I_D/I_G$ ratio is comparable to literature values reported for high-quality 9-AGNRs, supporting the structural integrity of the as-grown ribbons.[26–28] The observation of these characteristic signatures in pristine 9-AGNRs confirms successful synthesis and establishes a structural baseline for evaluating post-exposure changes.

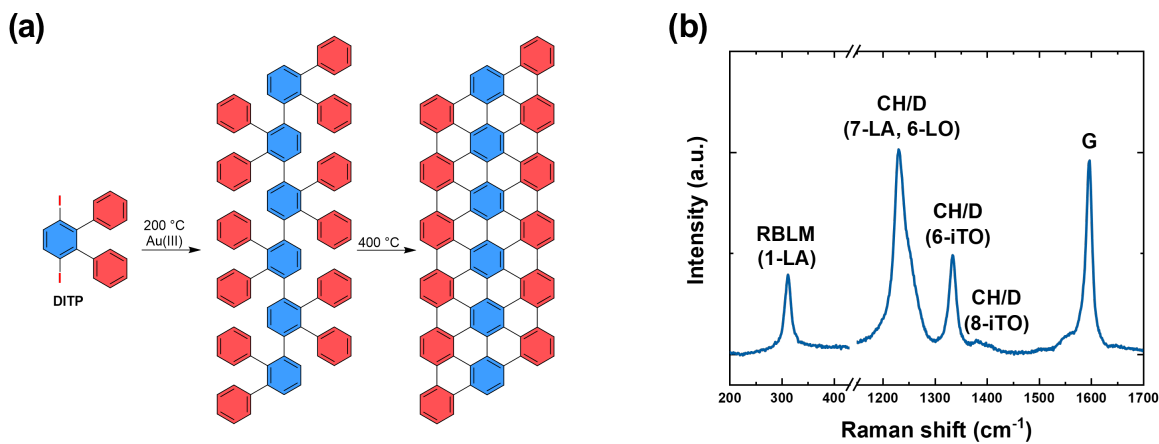

**Figure 1: (a)** Schematic illustration of the surface-assisted synthesis of a 9-AGNRs from the molecular precursor DITP on Au(111), including polymerization upon annealing (200 °C) and subsequent cyclodehydrogenation to form the fully conjugated GNR upon further annealing (400 °C). **(b)** Raman spectrum of the as-synthesized GNR, showing the characteristic RBLM, CH/D-related modes, and the G band.

As-grown 9-AGNRs were first transferred onto HfO$_2$/SiO$_2$/Si substrates, followed by gamma irradiation (see Methods section for details). The irradiated samples were subsequently characterized by Raman spectroscopy. Representative Raman spectra, normalized to the G-band intensity, are shown in **Figure 2a**. The average spectrum was derived from measurements of three individual samples, represented by the solid line, while individual raw data are displayed as semi-transparent lines to visualize sample-to-sample variation. We analyzed the irradiation-induced changes in terms of (i) spectral-shape evolution, (ii) disorder/defectiveness trends via $I_D/I_G$, (iii) peak-position shifts, and (iv) linewidths (FWHM).

The RBLM near ~396 cm$^{-1}$ is a width-sensitive vibrational mode and serves as a key fingerprint for verifying the formation of narrow GNRs.[20,27,29] The RBLM remained detectable after gamma irradiation (red curve in **Figure 2a**). This observation suggests that the ribbons largely retain their overall nanoribbon structures and do not undergo width-changing processes (e.g., extensive cutting or fusion). The CH/D peaks in the 1100-1500 cm$^{-1}$ region and G band near 1600 cm$^{-1}$ became broader, as discussed below, indicating the introduction of structural disorder.

After gamma irradiation, the $I_D/I_G$ ratio, a standard metric for disorder/defect density in carbon materials, showed a substantial increase (**Figure 2b**). The as-fabricated samples exhibit $I_D/I_G$ ~0.51. Because quantum confinement in 9-AGNRs can yield an observable D band even in defect-free GNRs, this baseline value can be regarded as a practical "defect-free" reference for this system rather than an indication of substantial disorder. After irradiation, $I_D/I_G$ increased to ~0.60, corresponding to an ~18% increase. This trend indicates an increase in defects induced by irradiation.

**Figure 2c** summarizes the peak-position changes of the major Raman modes. The RBLM peak exhibits only a slight redshift. Given that the RBLM frequency is highly sensitive to ribbon width and is known to shift significantly with a change of even one carbon atom in width, the observed small shift indicates that the ribbon width is largely preserved after irradiation. A comparable redshift of the RBLM peak was reported by Ma et al. for 7-AGNRs subjected to oxidative treatment at 520 °C,[23] supporting the interpretation that the present spectral changes are consistent with an oxidation-like degradation pathway.

This trend is also consistent with a termination-driven shift of the quantization condition: when H termination is replaced by an oxidized termination such as OH, the quantized wavevectors shift, and **Figure S2** shows a redshift of the folded RBLM, in agreement with our observation. For other modes, **Figure S2** suggests that the termination-induced shift can be mode dependent, which may account for the fact that some peaks blueshift while others change only weakly.

We further analyzed the changes in FWHMs for each mode (**Figure 2d**). Because FWHM reflects phonon lifetime and crystalline order, an increase in defect concentration typically enhances phonon scattering and broadens Raman peaks. Experimentally, all modes except the RBLM (7-LA/6-LO, 6-iTO, 8-iTO, G) showed clearly larger FWHM values after gamma irradiation (red symbols) than before irradiation (gray symbols). These linewidth increases corroborate a reduction in crystalline order and an increase in defect density induced by irradiation. The comparatively small increase in the RBLM linewidth again indicates that the ribbon width and overall ribbon framework remain largely intact.

The Raman spectroscopy results clearly indicate that gamma irradiation induces structural and/or chemical modifications in the 7-AGNRs. Following irradiation, an enhancement of the D band and a broadening of the overall peak profiles were observed. These features reflect the generation of defects and alterations in the local bonding states. Such Raman signatures can arise from multiple damage mechanisms, including oxidation, radiation-induced chemical reactions, and ballistic atomic displacement (knock-on damage).

First, oxidative processes represent a primary candidate. Ma et al. demonstrated that the armchair edges of 7-AGNRs remain structurally stable up to approximately 430 °C,[23] reporting that a significant activation energy is required to cleave the H-terminated C–H bonds for reaction with oxygen. In contrast, the present study reveals Raman spectral changes like those observed in high-temperature oxidation, despite the gamma irradiation being conducted near room temperature. This suggests that reactive species generated by irradiation may promote oxidation via pathways distinct from thermal activation.

When gamma rays pass through air, secondary electrons are generated via processes such as Compton scattering.[30] These electrons ionize and excite $O_2$ and $H_2O$ molecules, initiating radiolysis. Consequently, highly oxidative species such as ozone ($O_3$), atomic oxygen (O), and hydroxyl radicals (·OH) can be formed. These species can abstract hydrogen atoms from relatively stable C–H bonds or adding to the carbon skeleton, thereby introducing oxygen-containing functional groups, such as hydroxyl, epoxy, and carbonyl groups, to the edge carbons.[31,32] The introduction of such oxygen functional groups is thought to induce local changes in bonding states, causing bond length elongation and lattice strain,[33,34] as well as altering lattice vibrational frequencies due to increased molecular mass. These effects are likely reflected in the Raman spectra as peak shifts and broadening.[15,17,35]

On the other hand, the possibility of knock-on damage caused by secondary electrons or high-energy charged particles associated with gamma irradiation cannot be ruled out. Photons with energies of 1.17 and 1.33 MeV from a gamma-ray source (e.g., cobalt-60) can generate secondary electrons via Compton scattering that possess energies exceeding the threshold electron energy (approximately 80-90 keV) required to overcome the displacement energy of carbon atoms in graphene (approximately 22 eV).[36] Generally, in the knock-on process, the ballistic removal of carbon atoms leads to the formation of vacancies or reconstructed defects. The resulting inhomogeneity in ribbon width is expected to cause the disappearance of the RBLM or significant frequency shifts. However, in this study, the RBLM remained clearly detectable even after irradiation, with only limited frequency changes observed (**Figure 2a** and **2c**). Considering the extreme sensitivity of the RBLM to ribbon width, while knock-on events are physically possible, it is unlikely that extensive carbon removal capable of significantly altering the ribbon width is the dominant mechanism.

Nevertheless, these results do not imply that knock-on processes are entirely absent. It is possible that local and minute atomic displacements occur near edges or pre-existing defects, creating chemically unstable reaction sites that facilitate subsequent oxidation or functionalization.

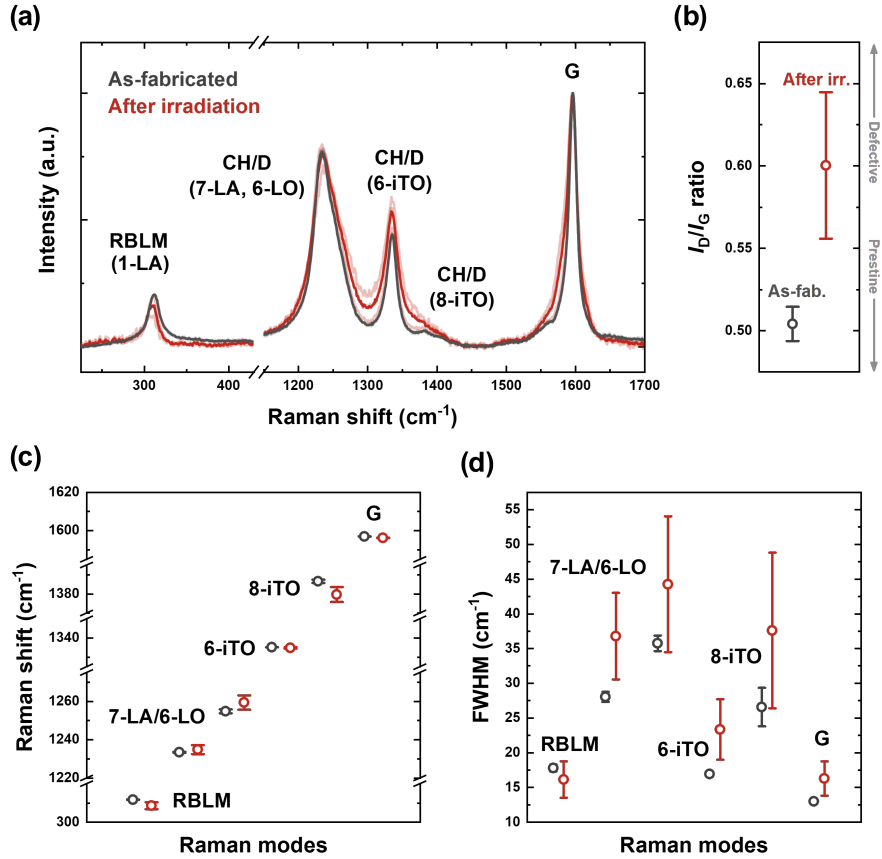

**Figure 2:** Raman characterization of 9-AGNRs before and after gamma irradiation. **(a)** Raman spectra of an as-fabricated sample (black) and after irradiation (red), with major modes labeled. **(b)** $I_D/I_G$ ratio extracted from Lorentzian peak fitting for the as-fabricated and irradiated states. **(c)** Peak center positions and **(d)** FWHM for the Raman modes obtained from the Lorentzian fits; gray and red symbols correspond to as-fabricated and irradiated states, respectively.

To investigate how the structural modification of the GNRs affects their electrical properties, FETs were fabricated using the architecture shown in **Figure 3a**, and the representative SEM image of the resulting devices are shown in **Figure 3b**. Detailed fabrication procedures and device parameters are provided in the Methods section. Transfer characteristics ($I_{DS}$-$V_{GS}$) were measured before and after gamma irradiation. (**Figure 3c**). The reported electrical characteristics are averaged over four devices, with individual data provided in the Supporting Information.

Electrical measurements obtained prior to irradiation show an on-current of $\sim 1.3 \times 10^{-9}$ A and an off-current of $\sim 5.6 \times 10^{-12}$ A, corresponding to an $I_{ON}/I_{OFF}$ ratio of $\sim 2.3 \times 10^2$. Analysis of the logarithmic transfer curve yields a subthreshold swing (SS) value of $\sim 1.38 \times 10^3$ mV/dec for the forward sweep and $1.24 \times 10^3$ mV/dec for the backward sweep. SS serves as a metric of gate-control efficiency, with lower values indicating stronger gate control. The transfer characteristics exhibit hysteresis, and SS was therefore extracted from the two sweep directions. Details of the SS extraction procedure are provided in the Supporting Information. Overall, these values are comparable to the previously reported results and provide a baseline for evaluating irradiation-induces changes.[26,37–39]

Following gamma irradiation, both the on-current and off-current decrease to ~2.7 × 10$^{-11}$ A and ~4.1 × 10$^{-12}$ A, respectively, yielding an $I_{ON}/I_{OFF}$ ratio of ~6.6 (a ~97% reduction compared with the pre-irradiation value). The SS significantly increases to ~3.2 × 10$^3$ mV/dec for the forward sweep and 2.4 × 10$^3$ mV/dec for the backward sweep, indicating a substantial degradation of gate-control efficiency after irradiation.

While the Raman spectra exhibit subtle changes after gamma irradiation, the device performance degrades drastically. This observation implies that factors beyond the slight Raman-detectable changes may contribute significantly to the severe device degradation. Furthermore, the magnitude of degradation appears substantially larger than that reported for gamma irradiated CNTFETs,[40] motivating a closer examination of why GNRFETs exhibit such high sensitivity to gamma rays and what differentiates their response from other carbon-based channels.

Aging is a possible source of device degradation. To rule out this contribution, we performed electrical measurements on the same devices at three time points: (1) immediately after fabrication (as-fabricated), (2) after storage in ambient air for one month, and (3) following gamma irradiation. The as-fabricated devices and those stored for one month exhibited negligible changes in the maximum on-state current, indicating that the FET architecture remains stable under ambient conditions over the ~1-month shipping/handling interval associated with the gamma irradiation experiment (see the Supporting Information). Therefore, ambient aging alone cannot account for the severe performance degradation observed after gamma-ray exposure.

Degradation of the gate dielectric is another possible origin of the observed performance change. To evaluate this possibility, we measured the gate leakage current before and after gamma irradiation. Because irradiation-induced defects in the dielectric can create additional conduction pathways, changes in gate leakage current provide a sensitive indicator of dielectric damage. As shown in the Supporting Information, the gate leakage current remains essentially unchanged after irradiation, suggesting that gamma irradiation induces negligible degradation of the gate dielectrics.

Bite defects, missing benzene rings at the ribbon edge, are a common form of structural disorder in GNRs.[13,26] Such defects could, in principle, be generated by knock-on displacement of carbon atoms under gamma irradiation. Although our Raman spectra indicate that the density of newly introduced bite defects does not increase appreciably, prior work has shown that even a single bite defect can strongly suppress the conductance of 9AGNRs.[13,26,38,41]

To assess whether bite defects could quantitatively account for the observed current decrease (i.e., the on-current drops to ~1% of the pre-irradiation value), we adopt a simple series-resistance model in which each bite defect introduces an additional local resistance ($\Delta R$), and these contributions add along the ribbon.[13] Using the reported single-defect suppression ($I_1/I_0 \sim 1/5$ at a fixed $V_{DS}$) as a calibration gives $R_1 = 5R_0$ and thus $\Delta R = R_1 - R_0 = 4R_0$; for $N$ defects, $R_N \sim R_0 + N\Delta R = R_0(1 + 4N)$, leading to $I_N/I_0 \sim 1/(1 + 4N)$. Here, we use the literature-reported defect density (~5 bite defects within a 20 nm-long 9-AGNR) as the baseline condition and denote the corresponding current as $I_5$.[26] Under this model, reaching $I_N/I_5 \sim 0.01$ requires $N \sim 525$ defects within 20 nm. This value far exceeds the number of unit cells in a 20 nm-long 9-AGNR (~47 unit cells), indicating that bite defects alone are unlikely to account for the severe current degradation.

Given that the severe current suppression is unlikely to be explained solely by a local-resistance increase from bite defects, we next consider oxidation-induced changes in the electronic properties suggested by the Raman analysis. Such changes could affect electrical transport without causing obvious structural damage to the GNR framework. Edge oxidation in 7-AGNRs has been reported and is known to modify the electronic structure, including band-gap modulation. In particular, oxidation reduces the band gap of 7-AGNRs from 2.6 eV to 2.3 eV and 1.9 eV for hydroxyl pair- and epoxy-terminated edges, respectively,[23] suggesting that a similar band-gap reduction may occur in 9-AGNRs. However, because oxidation in this context decreases (rather than increases) the band gap, a simple "bandgap opening" picture is unlikely to account for the pronounced loss of conductivity observed after gamma irradiation. Instead, oxidation may primarily introduce spatially nonuniform electronic perturbations along the ribbon.

Such spatially nonuniform electronic perturbations can have a profound impact on transport, particularly in low-dimensional systems. For example, it has been reported that the semimetal 1T'-WTe$_2$ undergoes a drastic transition from a metallic to an insulating state as its thickness is reduced from bulk to the few-layer regime.[42] This transition is attributed to the strong confinement of carriers within a nanoscale thickness, which amplifies the scattering effects of surface disorder induced by air exposure. In the presence of such disorder, the wave functions of scattered carriers undergo quantum interference and become localized, leading to a severe reduction in electrical conductivity, a phenomenon known as Anderson localization.

Because GNRs are quasi-1D materials, charge carriers are even more tightly confined than in 2D graphene or other 2D materials, making them significantly more sensitive to disorder. Previous studies on GNRs with widths of several to tens of nanometers have shown that even a relatively low density of edge defects can cause a marked suppression of conductance by Anderson localization.[43] Since the impact of this localization is known to scale inversely with GNR width,[43] the impact of disorder is expected to be even more pronounced in our ~1 nm-wide 9-AGNRs. This width-dependent sensitivity is also consistent with the observation that CNTs, which generally have larger effective diameters and fewer open edges than GNRs, exhibit relatively smaller degradation upon gamma-ray exposure.[40]

Taking all these factors into account, we conclude that the severe performance degradation observed after gamma irradiation is not primarily due to the destruction of the GNR carbon framework (as suggested by Raman) nor a simple uniform change in intrinsic electronic properties (such as bandgap opening). Instead, the dominant mechanism is likely the localization of carriers via quantum interference, driven by disorder likely introduced by edge oxidation (as suggested by Raman analysis). This disorder-induced localization is expected to be enhanced by the strong 1D confinement arising from the ultranarrow (~1 nm) width of 9-AGNRs, making electrical transport more sensitive to disorder than in 2D materials.

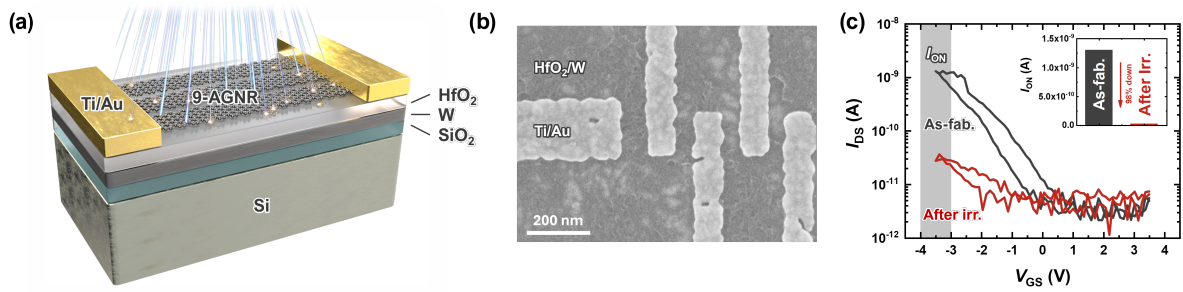

**Figure 3: (a)** Schematic illustration of the 9-AGNRFET architecture used to study the effects of gamma irradiation. **(b)** SEM image of the fabricated device (scale bar: 200 nm). **(c)** Transfer characteristics ($I_{DS}$–$V_{GS}$) measured before (as-fabricated) and after gamma irradiation, plotted with a logarithmic $I_{DS}$ axis in the main panel; the gray-shaded $V_{GS}$ window indicates the gate-voltage range used to extract $I_{ON}$. Inset: comparison of $I_{ON}$, defined as the drain current at $V_{GS}$ = -3.5 V, extracted from the gray-shaded $V_{GS}$ region before and after irradiation (linear $I_{ON}$ axis).

## Conclusions

We demonstrated a systematic evaluation of nine-atom-wide armchair graphene nanoribbon (9-AGNR) field-effect transistors under gamma irradiation. Raman spectroscopy confirmed that the nanoribbon framework and width remain largely intact, while enhanced D-band intensity, peak broadening, and minor shifts indicated a small increase of irradiation-induced disorder. Electrical measurements showed a dramatic reduction in on-state current and Ion/Ioff ratio, underscoring the extreme sensitivity of quasi-one-dimensional GNRs to subtle structural perturbations. Our analysis suggests that the degradation is more consistent with disorder-driven transport and carrier localization that are amplified by the quasi-one-dimensional confinement in 9-AGNRs than with catastrophic destruction of the GNR framework. These findings suggest that 9-AGNRFETs could serve as promising candidates for sensitive, nanoscale integrated sensors for monitoring high-energy radiation. This work provides a foundation for future studies that correlate chemical modifications with electrical transport behavior and that explore device-design strategies to improve radiation resilience and sensing performance.

## Experimental Methods

**Synthesis of Graphene Nanoribbons**

The 9-AGNRs were synthesized in a fully automated in-house ultra-high vacuum (UHV) system dedicated to GNR synthesis (Createc MiniMBE System Type RS2-M-4-FS). The precursor monomer 3,6'-di-iodine-1,1':2'1''-terphenyl (DITP) was sublimated onto Au/mica substrates (Phasis, Switzerland) that had been cleaned in a UHV chamber by $Ar^+$ sputtering and subsequent annealing. Detailed synthetic procedures for making DITP are provided in the Supporting Information.

**Transfer of Graphene Nanoribbons**

9-AGNRs were transferred to pre-patterned substrates using a conventional transfer method. In this process, the GNR/Au/mica was placed in an HCl solution, where the HCl attacks the interface between the Au and the mica. The released GNR/Au film floated on the surface of the solution and was then scooped up with the pre-patterned device substrate. After the GNR/Au film adhered to the device substrate, the Au layer was etched using a potassium iodide (KI) solution, leaving only the GNRs on the substrate.

**Raman Spectroscopy Characterization**

Raman spectroscopy of the 9-AGNRs was carried out on a Renishaw Raman microscope using a 785 nm laser. The laser power was maintained below 10 mW, and all measurements were taken with a 50× objective lens.

**Scanning Electron Microscopy Characterization**

SEM imaging was performed using a Hitachi S-4800 system with acceleration voltages of 15 and 20 kV.

**Preparation of Pre-Patterned Local Bottom Gate Chips**

Local gate structures were fabricated on heavily doped Si substrates capped with 100 nm of thermal $SiO_2$. An ~8 nm tungsten (W) layer was deposited by sputtering and subsequently patterned through standard photolithography, followed by selective wet etching in $H_2O_2$ to define the gate geometry. The gate dielectric, consisting of ~5.5 nm $HfO_2$, was deposited using atomic layer deposition at 135 °C. Photolithography and lift-off were then used to form alignment features and probing electrodes composed of a ~3 nm chromium (Cr) adhesion layer and a ~25 nm platinum (Pt) overlayer. After completing the metallization steps, the processed wafer was diced into individual chips, which served as the starting substrates for device assembly.

**Contact Metal Fabrication**

Source and drain contacts were defined using electron-beam lithography carried out on an Elionix ELS-7000 system, followed by metal deposition in a Temescal FC-2500 e-beam evaporator. After transferring the 9-AGNRs onto the pre-patterned substrates, a bilayer resist stack was applied for patterning. The bottom layer consisted of MMA EL6, spin-cast at 4000 rpm for 1 min and baked at 150 °C for 5 min, while the top layer of PMMA 950K A2 was deposited at 4500 rpm and baked at 180 °C for 10 min. Pattern writing was conducted using a 50 pA electron beam. After development, a Ti/Au (0.5/15 nm) metal was evaporated to form the source–drain electrodes. The fabricated devices exhibited channel lengths of ~70–35 nm, with electrode widths of 100–200 nm and lengths of 150-200 nm.

**Electrical Transport Measurements**

Electrical characterization was carried out on a Lakeshore TTPX cryogenic probe station equipped with an M81-SSM synchronous source-measure unit and controlled via MeasureLINK software. All device measurements were obtained under ambient conditions.

**XPS Characterization**

X-ray photoelectron spectroscopy measurements were performed using a Kratos Axis Ultra 165 spectrometer. The system was operated at a base pressure of $\leq 2 \times 10^{-8}$ Torr. Monochromatic Al K$\alpha$ radiation (hv = 1486.6 eV) was used as the excitation source, generated at 300 W (20 mA, 15 keV).

**Gamma Irradiation Methods**

The GNR devices were irradiated using a $^{60}$Co source at Sandia National Laboratory's Gamma Irradiation Facility (GIF). Samples were exposed to gamma irradiation with a dose rate of 41.63 rad(Si)/sec. Total irradiation times of 180 minutes, nearly 7 hours, and over 18 hours resulted in total accumulated gamma doses of 449.6 krad(Si) for the three sample conditions that are reported in this manuscript.

# Acknowledgments

This work was supported in part by NSF CHE-2235143 (GNR growth) and Semiconductor Research Corporation (LMD-3144.001) (device fabrication). Atomic layer deposition was performed at Molecular Foundry at Lawrence Berkeley National Laboratory. Raman spectroscopy and scanning electron microscopy were conducted at Kuiper–Arizona Laboratory for Astromaterials Analysis, and X-ray photoelectron spectroscopy measurements were carried out at the Laboratory for Electron Spectroscopy and Surface Analysis (LESSA) at University of Arizona. Gamma irradiation experiments were performed at Sandia National Laboratories. K.S.P. acknowledges partial support from University of Arizona's AzRISE program. Z.M. acknowledges partial support from Arizona Technology and Research Initiative Fund. B.G. and K.S.P. thank the facility staff at Sandia National Laboratories for technical support during gamma testing.

# Author Contributions

B.G. conceived the study on gamma irradiation of GNRs. B.G. and K.S.-P. performed the gamma irradiation experiments. J.T.N. supervised the synthesis of the molecular precursors used for GNR growth, which was carried out by J.G.F. and E.T.H. Z.M. supervised the overall study and led the efforts on synthesis, device integration, and pre- and post-irradiation electrical and spectroscopic characterization, which were conducted by M.Y., K.Y., and A.M.D. K.Y. led the manuscript preparation in close collaboration with M.Y., with input from all authors. All authors reviewed and approved the final manuscript.

*Kentaro Yumigeta[a,#], Muhammed Yusufoglu[a,#], John G. Federice[b], Elena T. Hughes[b], Ahmet Mert Degirmenci[c], Jon T. Njardarson[b], Kelly Simmons-Potter[a,c,d], Barrett G. Potter[a,d,*], and Zafer Mutlu[a,c,e,*]*

[a] *Department of Materials Science & Engineering, University of Arizona, Tucson, Arizona 85721, USA*

[b] *Department of Chemistry and Biochemistry, University of Arizona, Tucson, Arizona 85721, USA*

[c] *Department of Electrical and Computer Engineering, University of Arizona, Tucson, Arizona 85721, USA*

[d] *James C. Wyant College of Optical Sciences, University of Arizona, Tucson, Arizona 85721,*

[e] *Department of Physics, University of Arizona, Tucson, Arizona 85721, USA*

[#] *Kentaro Yumigeta and Muhammed Yusufoglu contributed equally to this study.*

*\* Corresponding Authors: zmutlu@arizona.edu, bgpotter@arizona.edu*

# Table of Contents

# S1. DITP Monomer Synthesis and Characterization

## S1.1 General Experimental Methods

Reactions were monitored by thin layer chromatography (TLC) carried out on Supelco 250μm silica gel 60-F254 plates. Plates were visualized using mainly UV lamp and KMnO$_4$. Flash Chromatography was done with SiliaFlash® F60 (particle size 40-63μm). $^1$H, $^{13}$C NMR data was acquired on Bruker NEO 500 MHz NMR instrument. The spectra were referenced using residual solvent as internal reference for $^1$H and $^{13}$C NMR (CDCl$_3$: 7.26 ppm for $^1$H NMR, 77.16 ppm for $^{13}$C NMR). Signals are reported as follows: s (singlet), d (doublet), t (triplet), q (quartet), dd (doublet of doublets), dt (doublet of triplets), dq (doublet of quartets), dtt (doublet of triplets of triplets), ddq (doublet of doublets of quartets), br s (broad singlet), m (multiplet). Coupling constants are reported in hertz (Hz).

All reactions were performed under an argon atmosphere with dry solvents unless otherwise stated. Dry tetrahydrofuran (THF), dichloromethane (DCM), and dimethylformamide (DMF) were obtained by passing previously degassed solvents through activated alumina columns.

1,2-dibromobenzene was purchased from AA Blocks. Chlorotrimethylsilane (TMSCl) was purchased from Thermo Scientific. S Phos Pd G4 was purchased from Ambeed. Phenylboronic acid was purchased from Oakwood Chemical. Potassium phosphate tribasic (K$_3$PO$_4$) was purchased from AA Blocks. Iodine monochloride (ICl) was purchased from Thermo Scientific.

## S1.2 Experimental Procedures

**Overall Synthetic Scheme for Accessing DITP (4) from 1,2-Dibromobenzene:**

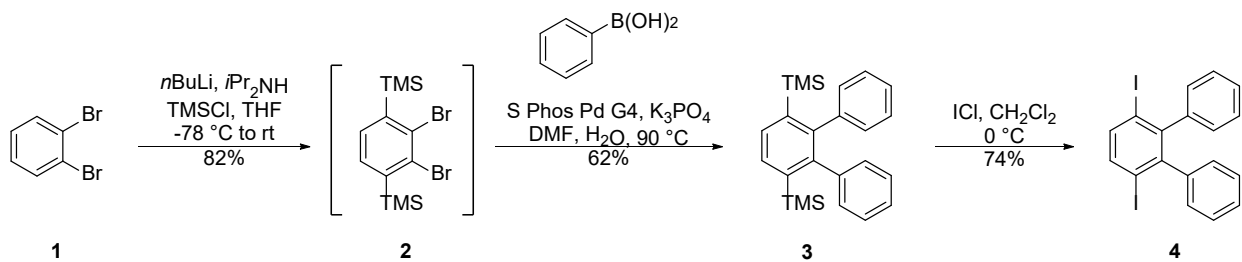

**Synthesis of (2,3-dibromo-1,4-phenylene)bis(trimethylsilane) (2)**

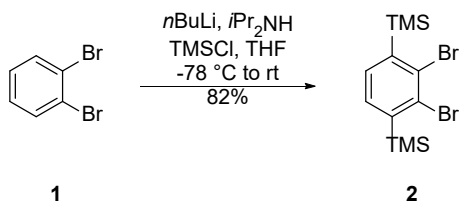

Following a modified procedure.[1] To a flame-dried 20 mL vial was added dry tetrahydrofuran (THF) (6.9 mL) and freshly distilled $i$Pr$_2$NH (1.0 mL, 7.20 mmol, 2.4 eq) under an argon atmosphere. *N*-BuLi (2.42 M in hexanes, 2.9 mL, 6.90 mmol, 2.3 eq) was added at 0 °C and the solution stirred for at least 10 minutes prior to use.

To a separate flame-dried 25 mL round bottom flask was added 1,2-dibromobenzene **1** (0.36 mL, 3.00 mmol, 1.0 eq), freshly distilled chlorotrimethylsilane (TMSCl) (0.84 mL, 6.60 mmol, 2.2 eq) and dry THF (5.0 mL, 0.6 M relative to **1**) The flask was cooled to -78 °C using a dry ice acetone bath and the LDA solution was added dropwise. This stirred at -78 °C for 6 hours and was then allowed to slowly warm to room temperature over 18 hours. The reaction was quenched with HCl (1 M in H$_2$O, 7.2 mL, 7.20 mmol, 2.4 eq). The reaction was diluted with H$_2$O and diethyl ether (Et$_2$O), transferred to a separatory funnel and extracted three times with Et$_2$O. The resulting organic layer was washed with brine and dried over magnesium sulfate (MgSO$_4$) which was then filtered and concentrated. The resulting crude mixture was then purified via column chromatography (hexanes) to afford (2,3-dibromo-1,4-phenylene)bis(trimethylsilane) **2** as a gel-like colorless solid (931 mg, 2.45 mmol, 82%). Due to how non-polar the product is, a minor impurity that is inseparable by column chromatography co-elutes with the product. The semi-pure product is carried forward to the next step without further purification.

**Note:** All spectral data was in accordance with the literature.

**$^1$H NMR (500 MHz, CDCl$_3$)** δ 7.33 (s, 2H), 0.39 (s, 18H).
**$^{13}$C NMR (126 MHz, CDCl$_3$)** δ 145.9, 134.1, 133.5, -0.3.

### Synthesis of 3',6'-bis(trimethylsilyl)-1,1':2',1''-terphenyl (3)

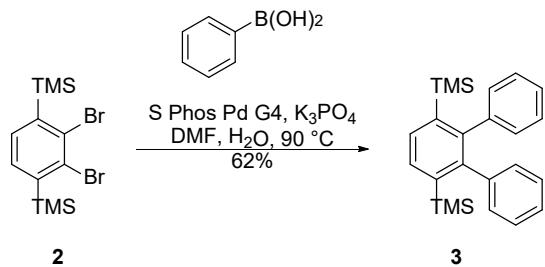

To a flame dried 50 mL round bottom flask was added (2,3-dibromo-1,4-phenylene)bis(trimethylsilane) **2** (1.39 g, 3.65 mmol, 1.0 eq), phenyl boronic acid (913 mg, 7.49 mmol, 2.05 eq), S Phos Pd G4 (290 mg, 0.37 mmol, 0.10 eq), and K$_3$PO$_4$ (4.65 g, 21.91 mmol, 6.0 eq) under an argon atmosphere. Dimethylformamide (DMF) (14.6 mL) and argon sparged H$_2$O (3.6 mL) (18.2 mL, 0.2 M relative to **2**) were added and the mixture was heated to 90 ºC for 16-24 hours. After cooling to room temperature, the mixture was diluted with H$_2$O, transferred to a separatory funnel, and extracted three times with ethyl acetate (EtOAc). The combined organic layers were washed three times with H$_2$O, followed by brine. The resulting organic layer was dried over Na$_2$SO$_4$ which was then filtered and concentrated. The resulting crude mixture was then purified via column chromatography (hexanes) to afford 3',6'-bis(trimethylsilyl)-1,1':2',1''-terphenyl **3** as a white solid (850 mg, 2.27 mmol, 62%).

**Note:** All spectral data was in accordance with the literature.

**$^1$H NMR (500 MHz, CDCl$_3$)** δ 7.67 (s, 2H), 7.11 – 7.05 (m, 6H), 7.00 – 6.96 (m, 4H), -0.04 (s, 18H).
**$^{13}$C NMR (126 MHz, CDCl$_3$)** δ 147.3, 142.4, 140.1, 132.8, 131.1, 126.9, 126.3, 0.6.

**Synthesis of 3',6'-diiodo-1,1':2',1''-terphenyl (4) – (DITP):**

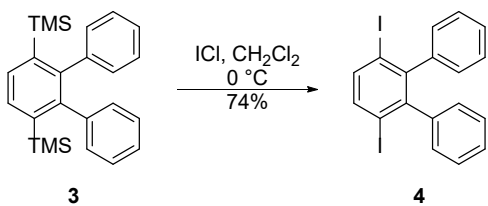

Following a modified procedure[1], to a flame dried 25 mL pear-shaped flask was added iodine monochloride (ICl) (1.10 g, 6.77 mmol) under an argon atmosphere. Dry DCM (13.5 mL; 0.5 M relative to ICl) was added and swirled until all the ICl dissolved.

In a separate 100 mL round bottom flask was added 3',6'-bis(trimethylsilyl)-1,1':2',1''-terphenyl **3** (850 mg, 2.27 mmol, 1.0 eq) under an argon atmosphere. Dry DCM (38 mL, 0.06 M relative to **3**) was added and the solution was cooled to 0 °C. ICl (0.5 M in DCM, 11.8 mL, 5.90 mmol, 2.6 eq) was added dropwise. The reaction was monitored by TLC and once starting material had been consumed (typically 1 hour), the reaction was quenched with 30% Na$_2$S$_2$O$_3$ (double the volume of DCM) and transferred to a separatory funnel. The mixture was extracted three times with DCM and the combined organic layers were washed with brine. The mixture was dried over Na$_2$SO$_4$, filtered, and concentrated. The resulting crude mixture was then purified via recrystallization (EtOH) to afford 3',6'-diiodo-1,1':2',1''-terphenyl **4 (DITP)** as white needles (810 mg, 1.68 mmol, 74%).

**Note:** The reaction was conducted under light exclusion using foil.

**$^1$H NMR (500 MHz, CDCl$_3$)** δ 7.64 (s, 2H), 7.20 – 7.08 (m, 6H), 6.96 – 6.92 (m, 4H).
**$^{13}$C NMR (126 MHz, CDCl$_3$)** δ 146.9, 144.1, 139.5, 129.7, 127.6, 127.3, 100.9.

## S1.3 NMR Characterization

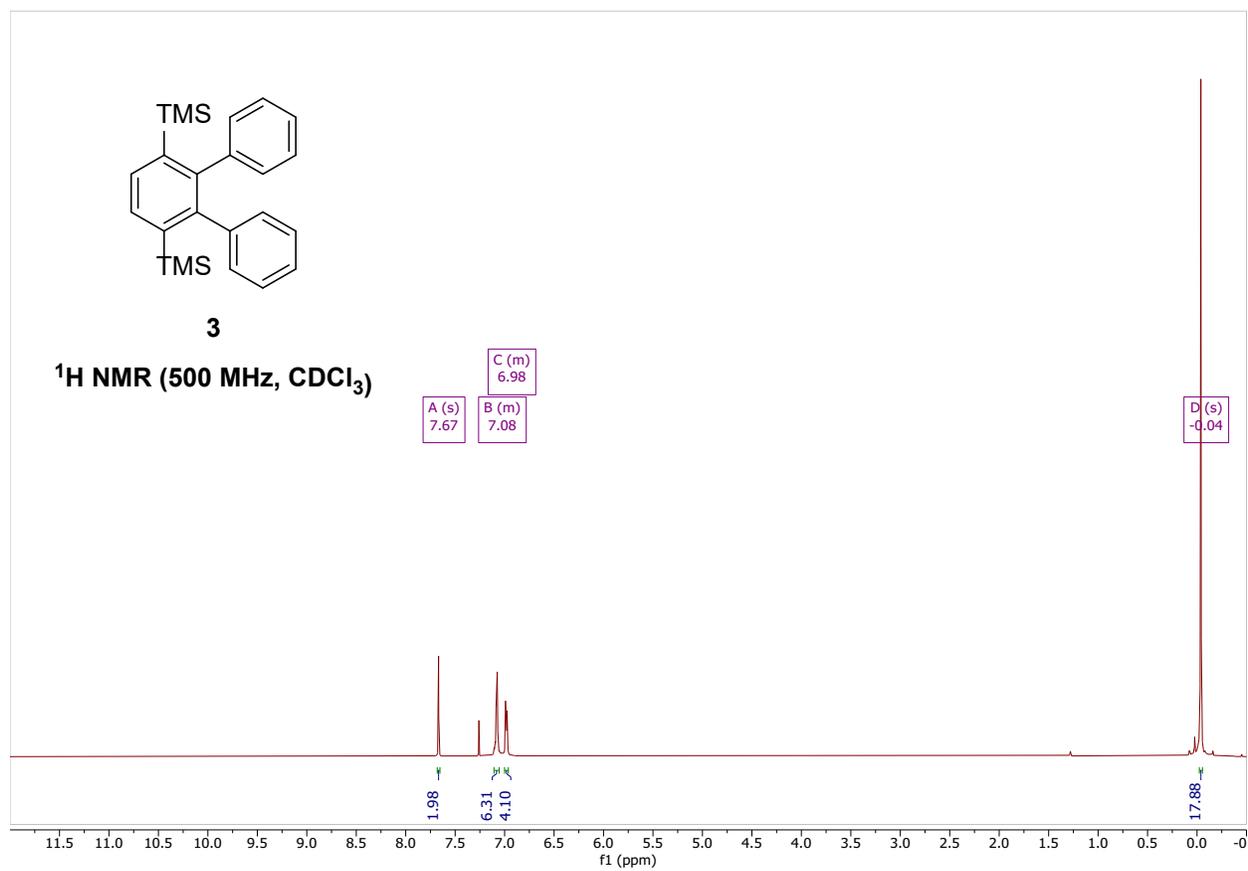

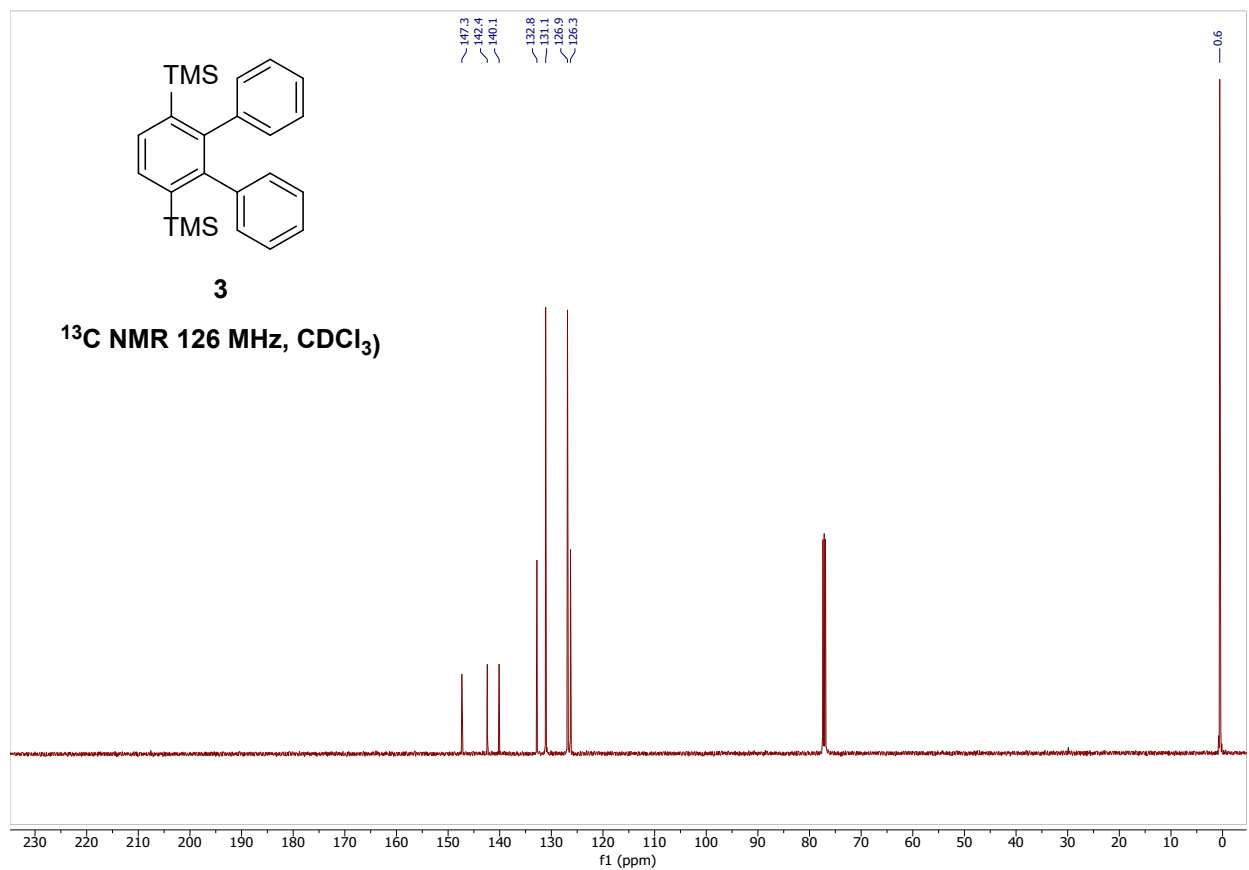

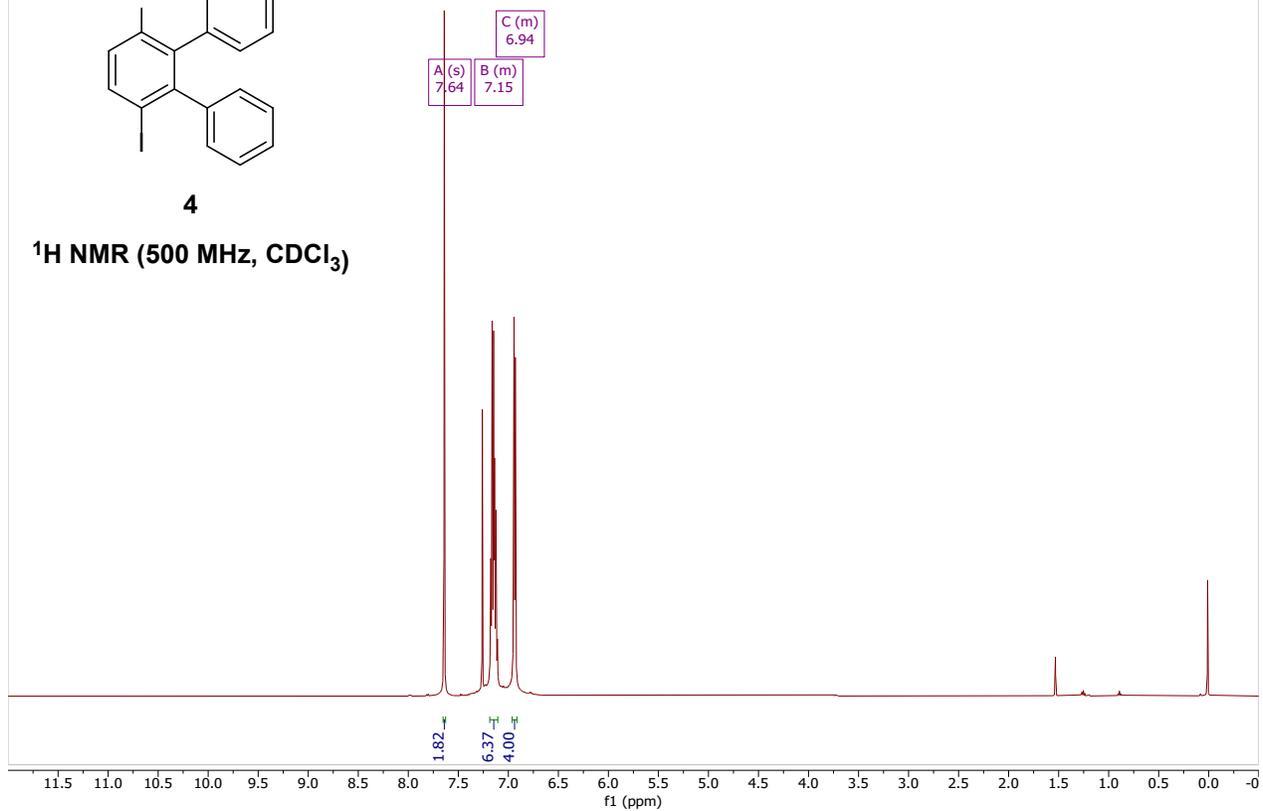

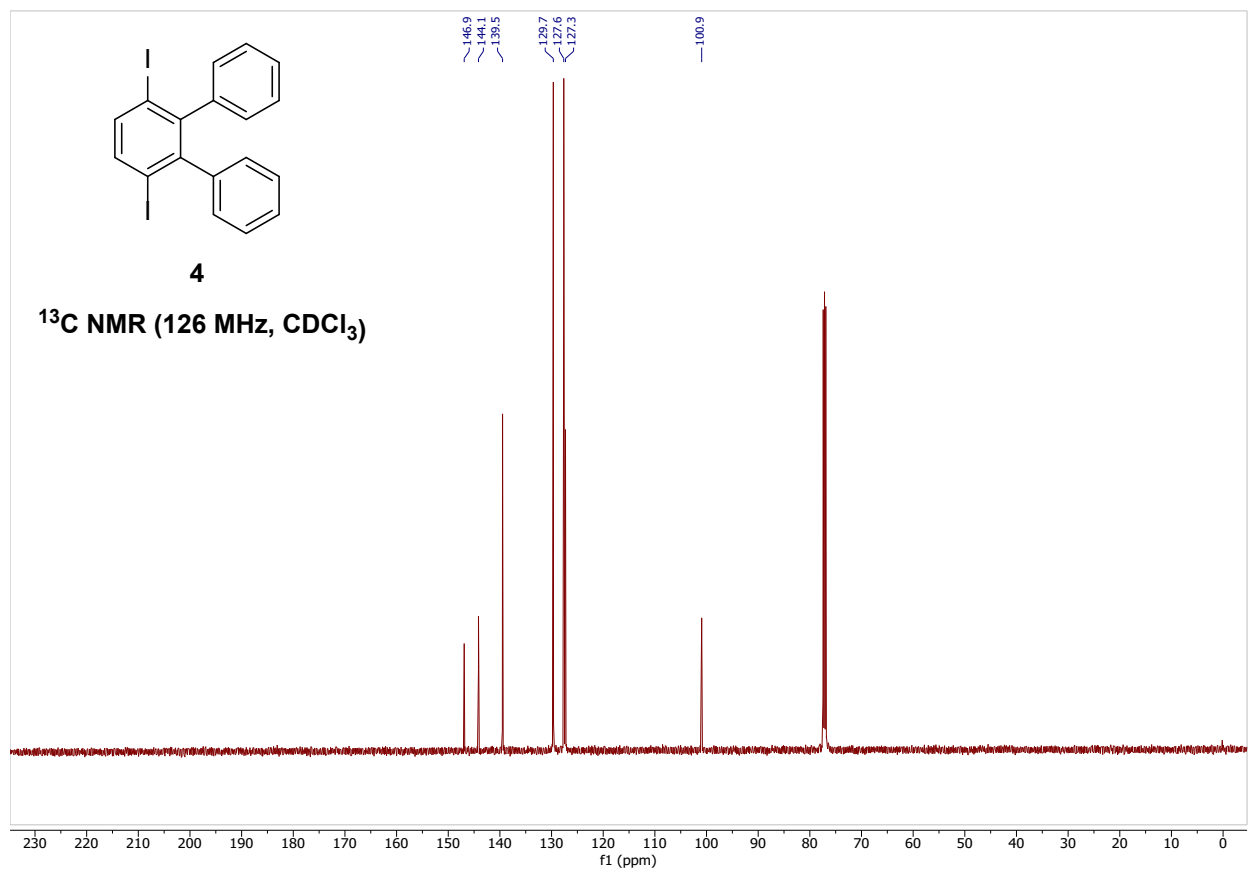

## S2. Raman Spectroscopy Characterization and Data Analysis

### S.2.1 Raman Spectra of 9-AGNRs (As-grown and transferred)

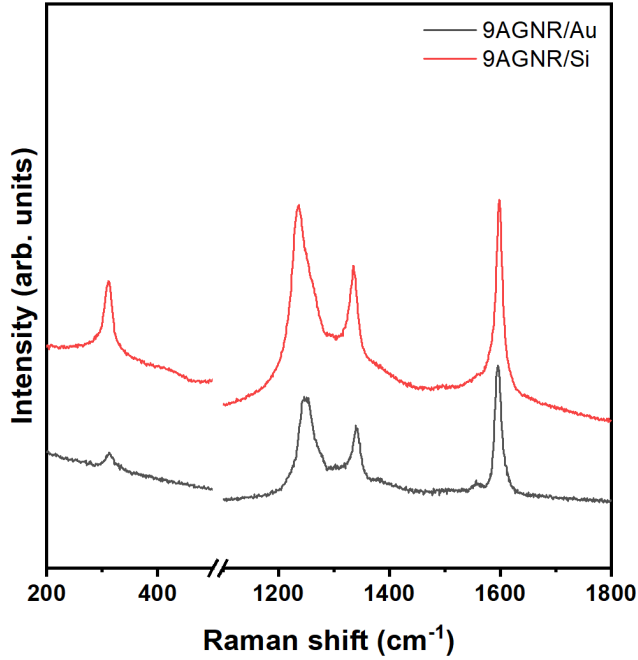

**Figure S1**: Raman spectra of 9-AGNRs on Au/mica and after transfer onto the pre-patterned device.

### S2.2 Phonon Dispersion of Graphene and Zone-Folding

Raman-active vibrational features in AGNRs are assigned via zone folding: the 1D Brillouin zone (BZ) is mapped onto the 2D graphene BZ, and the graphene phonon dispersion is sampled at quantized GNR wavevectors. The finite ribbon width quantizes the phonon-wavevector component perpendicular to the ribbon axis.[2,3] The boundary condition can be written as

$$k_{\perp,n} W_{\text{eff}} = n\pi, (n = 0, 1, \ldots, N-1)$$

where $k_{\perp,n}$ are the allowed width-direction wavevectors (perpendicular to the ribbon axis) and $W_{\text{eff}}$ is the effective ribbon width. Here, $N$ corresponds to the $N$ in $N$-AGNR (the dimer-line index). This expression follows the AGNR quantization condition, and we account for edge-termination effects by introducing an effective width,[4]

$$W_{\text{eff}} = W + \delta W.$$

Here, $W$ denotes the geometric AGNR width of the carbon backbone ($W = \frac{1}{2}(N-1)a_0$ with $a_0 = 2.46$ Å), and $\delta W$ captures the termination-dependent modification of the effective width arising from changes in the edge-region mass distribution and local elastic response. We treat $\delta W$ as a termination-specific correction: $\delta W = 1.8$ Å for H-terminated edges, whereas $\delta W = 5.5$ Å for OH-terminated edges.[4] H termination is commonly used to represent relatively intact GNR edges, whereas OH termination can occur after oxidative modification.[5]

In reciprocal space, each allowed $k_{\perp,n}$ defines a "cutting line" in the graphene BZ; the set of cutting lines is separated by $\Delta k_\perp = \pi/W_{\text{eff}}$. Ribbon phonon frequencies are obtained by sampling graphene dispersion at cutting-line intersections and folding the sampled points to the GNR Γ point (zone center).

Raman activity of folded modes depends on whether they transform as Raman-allowed irreducible representations at the GNR Γ point.[6,7] Acoustic-branch modes are Raman-active for odd zone-folding order (odd $n$), whereas optical-branch modes become Raman active for even zone-folding order (even $n$). For acoustic branches, the fundamental acoustic modes are not Raman active in nanoribbons. The first Raman-allowed folded graphene-LA mode is the radial breathing-like mode (RBLM). In the zone-folding construction, the quantized wavevector points across the ribbon width, so the width-breathing displacement is parallel to the quantized wavevector and thus follows the graphene-LA branch. **Figure S2** illustrates the termination dependence of the zone-folding construction: because OH termination increases $W_{\text{eff}}$ more strongly ($\delta W = 5.5$ Å vs 1.8 Å), the cutting-line spacing is reduced and the set of folded phonon wavevectors shifts accordingly.

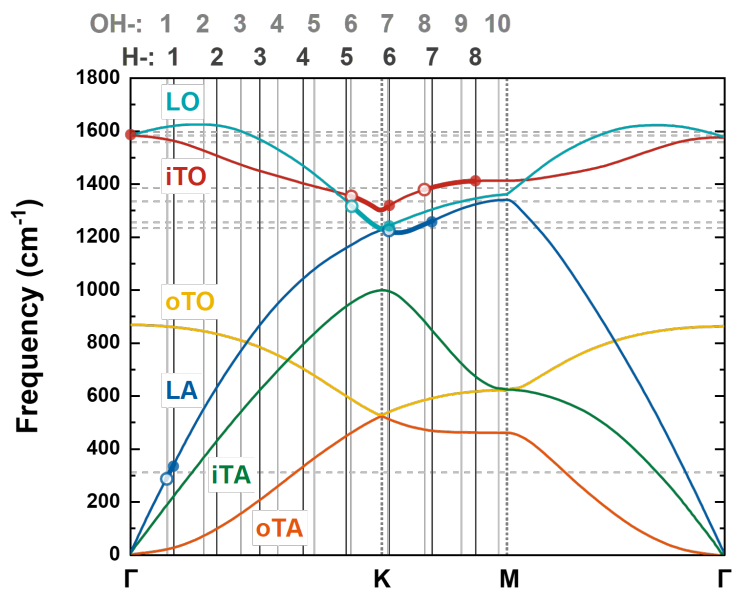

**Figure S2:** Phonon dispersion of graphene. The gray lines represent the Brillouin-zone cutting lines arising from quantization of the phonon wavevector in the width direction of the 9-AGNR. Gray dashed horizontal lines mark the peak positions used for Lorentzian fitting of the experimental spectra. Intersections between the cutting lines and the phonon dispersion indicate candidate folded phonon modes and are highlighted by circles: filled circles correspond to the pristine (H-terminated) case, whereas open circles correspond to the oxidized (OH-terminated) case. Adapted from Ref[8].

## S2.3 Lorentzian Peak Fitting Procedure and Fitted Parameters

The spectral peaks were fitted with a Lorentzian line shape to extract the peak parameters. The tables below report the best-fit values for the peak center position, full width at half maximum, and integrated area. The values in parentheses represent the standard errors derived from the fitting procedure, expressed in terms of the least significant digits (e.g., 1570.1(22) corresponds to 1570.1 ± 2.2).

**Table S1:** Summary of the Lorentzian fitting parameters for the Raman spectra of the as-fabricated graphene nanoribbon devices. The standard errors of the fits are indicated in parentheses.

| | | Raman shift (cm$^{-1}$) | | | FWHM (cm$^{-1}$) | | | Area | | |
|---|---|---|---|---|---|---|---|---|---|---|
| | | Device #1 | #2 | #3 | Device #1 | #2 | #3 | Device #1 | #2 | #3 |
| RBLM | | 311.654(69) | 311.878(74) | 311.685(76) | 17.52(23) | 17.51(24) | 18.38(25) | 1.553(16) | 1.499(17) | 1.757(20) |
| CH/D | 7-LA, 6-LO | 1233.61(15) | 1233.69(11) | 1233.14(13) | 28.01(38) | 28.79(29) | 27.33(34) | 8.45(22) | 8.66(16) | 8.73(22) |
| | | 1254.76(52) | 1255.90(41) | 1253.86(46) | 35.8(10) | 34.61(86) | 36.85(84) | 4.49(24) | 3.87(17) | 5.27(24) |
| | 6-iTO | 1335.220(68) | 1335.216(59) | 1335.260(59) | 16.67(22) | 17.34(19) | 16.82(19) | 3.194(32) | 3.196(27) | 3.490(31) |
| | 8-iTO | 1385.8(12) | 1386.49(90) | 1387.55(87) | 23.7(40) | 26.9(30) | 29.2(29) | 0.299(41) | 0.405(37) | 0.544(44) |
| G-like (low frequency) | | 1557.28(88) | 1556.61(84) | 1557.80(94) | 24.2(26) | 24.5(27) | 24.5(29) | 0.457(58) | 0.441(60) | 0.499(70) |
| G-like (shoulder) | | 1583.50(82) | 1584.9(11) | 1584.08(94) | 20.4(30) | 21.7(32) | 19.5(34) | 0.64(13) | 0.75(18) | 0.63(16) |
| G1 | | 1596.952(32) | 1597.036(39) | 1596.979(36) | 12.93(11) | 13.02(15) | 13.06(12) | 5.805(85) | 5.52(12) | 6.49(10) |

**Table S2:** Summary of the Lorentzian fitting parameters for the Raman spectra of the devices after gamma irradiation. The standard errors of the fits are indicated in parentheses.

| | | Raman shift (cm$^{-1}$) | | | FWHM (cm$^{-1}$) | | | Area | | |
|---|---|---|---|---|---|---|---|---|---|---|
| | | Device #1 | #2 | #3 | Device #1 | #2 | #3 | Device #1 | #2 | #3 |
| RBLM | | 308.93(17) | 306.66(35) | 310.236(66) | 16.22(54) | 18.7(11) | 13.49(21) | 3.84(10) | 3.97(20) | 15.88(19) |
| CH/D | 7-LA, 6-LO | 1234.35(22) | 1237.43(35) | 1232.79(11) | 34.73(54) | 43.78(78) | 31.85(29) | 34.2(12) | 53.9(26) | 136.8(21) |
| | | 1257.69(82) | 1263.7(15) | 1256.86(36) | 45.4(14) | 53.4(28) | 33.97(83) | 19.4(14) | 23.2(30) | 52.2(21) |
| | 6-iTO | 1334.875(75) | 1335.313(91) | 1334.513(60) | 23.35(28) | 27.70(38) | 18.99(20) | 17.84(20) | 29.31(45) | 49.87(41) |
| | 8-iTO | 1376.59(98) | 1378.5(13) | 1384.03(81) | 33.8(34) | 50.2(40) | 28.8(27) | 2.50(23) | 6.43(55) | 6.97(55) |
| G-like (low frequency) | | 1556.6(22) | 1554.5(29) | 1558.21(83) | 34.8(49) | 29.3(77) | 26.5(23) | 2.57(60) | 2.15(91) | 7.35(86) |
| G-like (shoulder) | | 1581.02(50) | 1580.56(69) | 1582.30(51) | 22.6(25) | 25.2(30) | 19.8(23) | 6.0(10) | 11.6(22) | 10.0(16) |
| G1 | | 1596.130(64) | 1596.30(12) | 1596.274(29) | 16.35(20) | 18.72(35) | 13.789(97) | 23.77(55) | 34.7(14) | 81.02(96) |

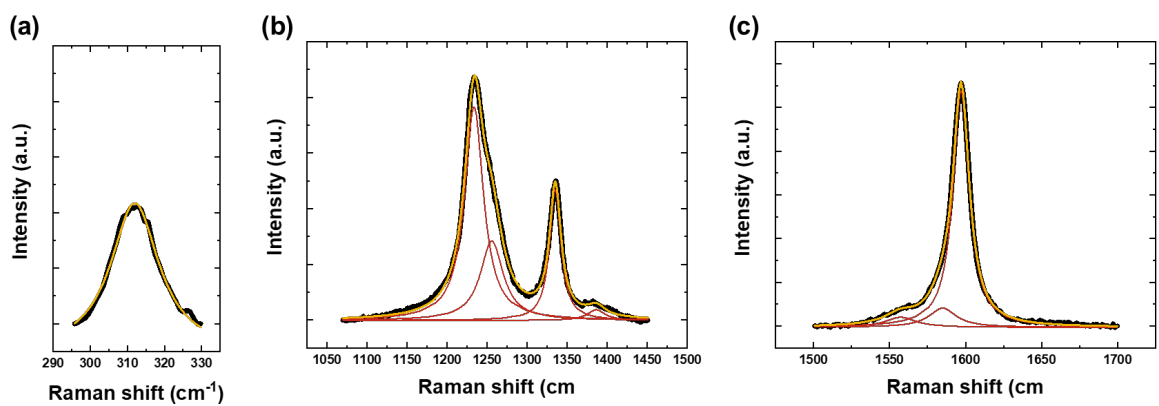

**Figure S3:** Representative Lorentzian peak fitting of Raman spectra in the **(a)** RBLM, **(b)** CH/D, and **(c)** G peak regions. Black curves show the experimental data, red curves show the individual fitted peaks, and the yellow curve shows the summed fit.

# S3. Device Fabrication and Structural Characterization

## S3.1 SEM imaging of Fabricated GNRFET Devices

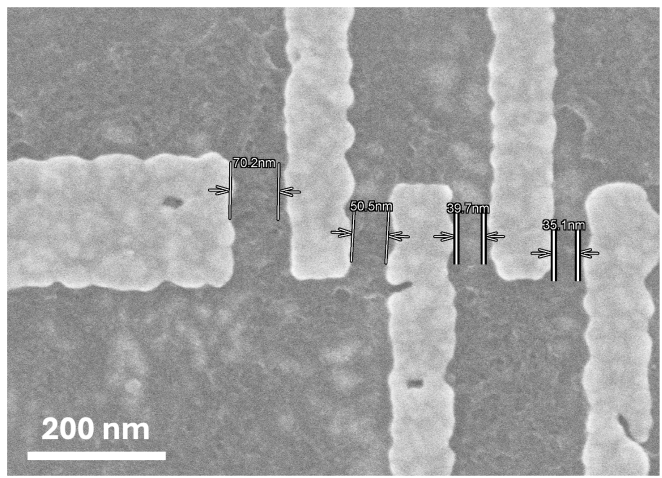

**Figure S4:** SEM image of a GNRFET showing channel lengths for four different electrode pairs, varying from approximately 70 to 35 nm.

## S3.2 Optical Imaging of Large-Area Device Arrays

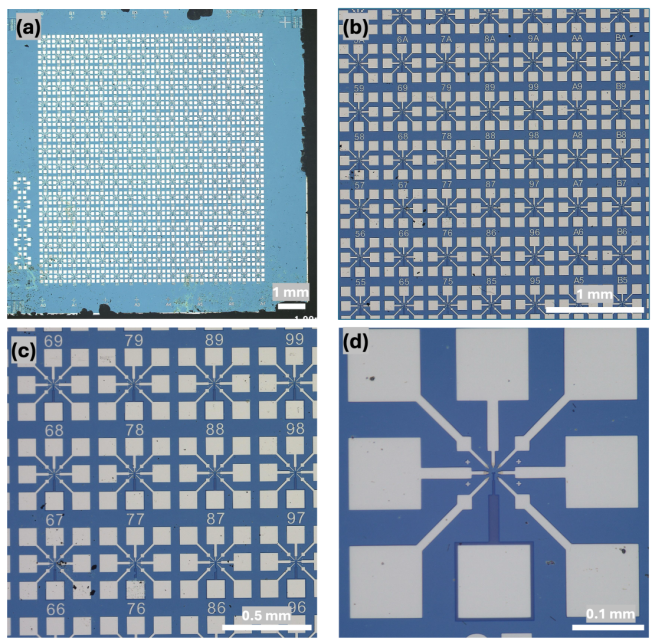

**Figure S5**: Optical images of the chip containing over 1000 9-AGNR transistors at different magnifications.

# S4. Electrical Transport Characterization

## S4.1 Gate Leakage and Measurement Stability

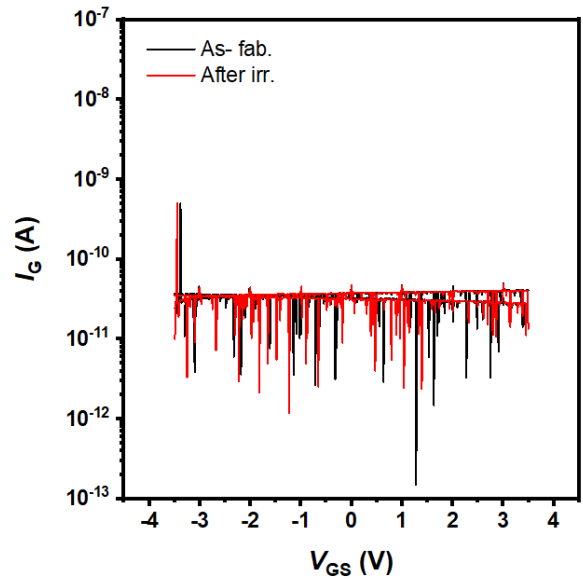

**Figure S6:** Leakage current ($I_G$) before and after irradiation showing no observable change.

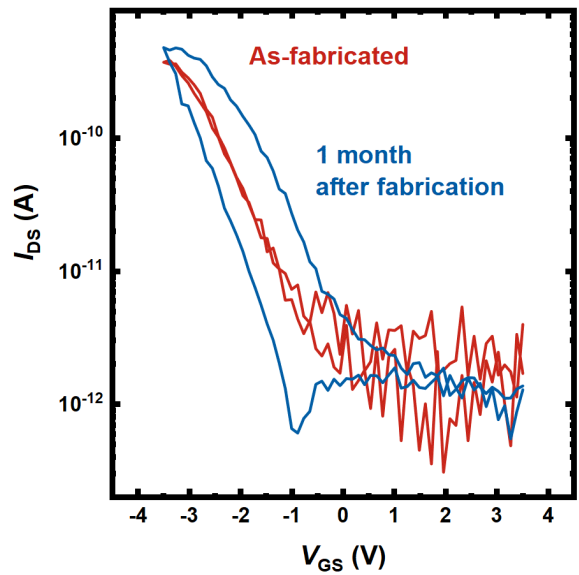

**Figure S7:** Transfer characteristics of the 9-AGNRFET device measured immediately after fabrication and again one month later, showing no significant change in electrical performance.

## S4.2 Transfer Characteristics ($I_{DS}$-$V_{GS}$) Before and After Irradiation

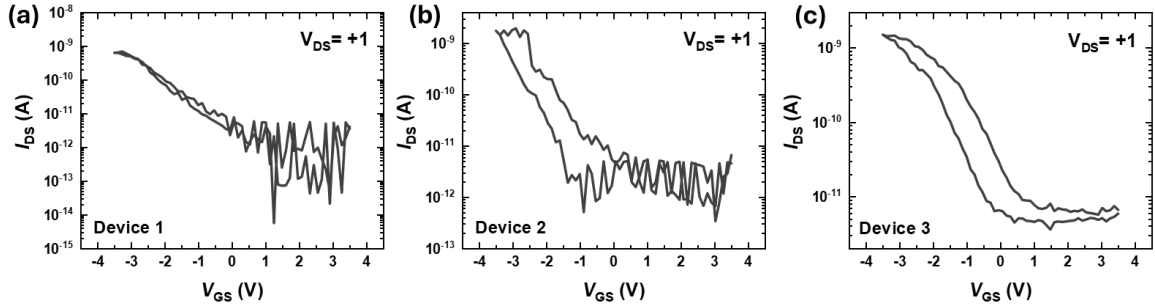

**Figure S8:** Transfer characteristics ($I_{DS}$–$V_{GS}$) of three GNRFET devices used to construct the averaged transfer curve prior to irradiation. All devices were measured under identical biasing conditions at room temperature and ambient pressure.

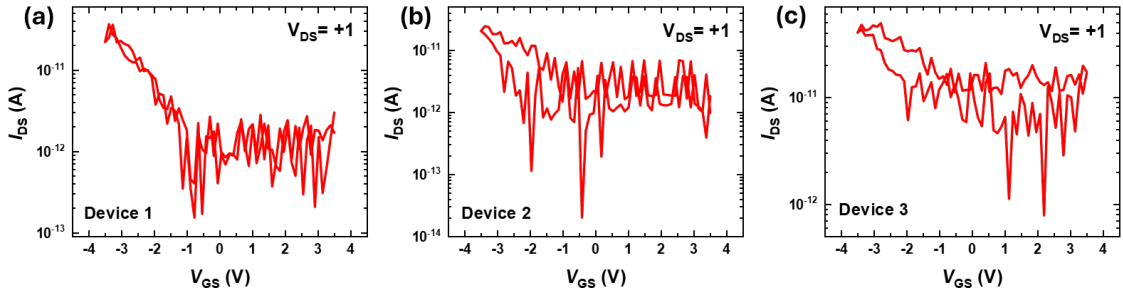

**Figure S9:** Transfer characteristics ($I_{DS}$–$V_{GS}$) of three GNRFET devices used to construct the averaged transfer curve after irradiation. All devices were measured under identical biasing conditions at room temperature and ambient pressure.

## S4.3 Output Characteristics ($I_{DS}$-$V_{DS}$) Before and After Irradiation

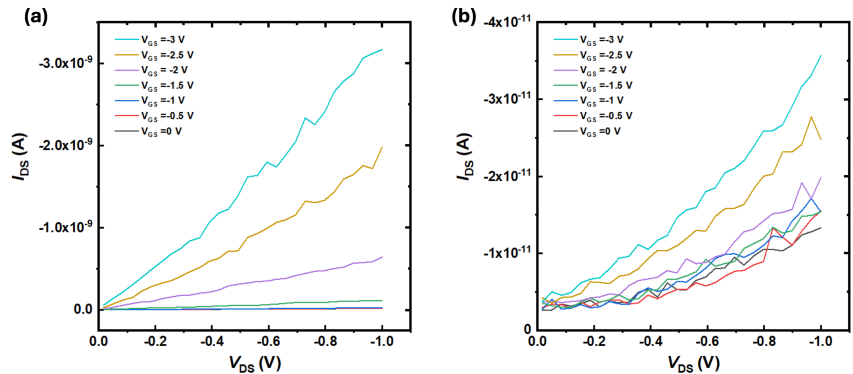

**Figure S10:** Average output characteristics ($I_{DS}$–$V_{DS}$) of a GNRFET device set measured before and after irradiation. Panel (a) shows the $I_{DS}$–$V_{DS}$ curves averaged over three devices prior to irradiation. Panel (b) presents the corresponding curves re-measured on the same devices after irradiation under the same conditions.

# S5. Subthreshold Swing (SS) Extraction and Analysis

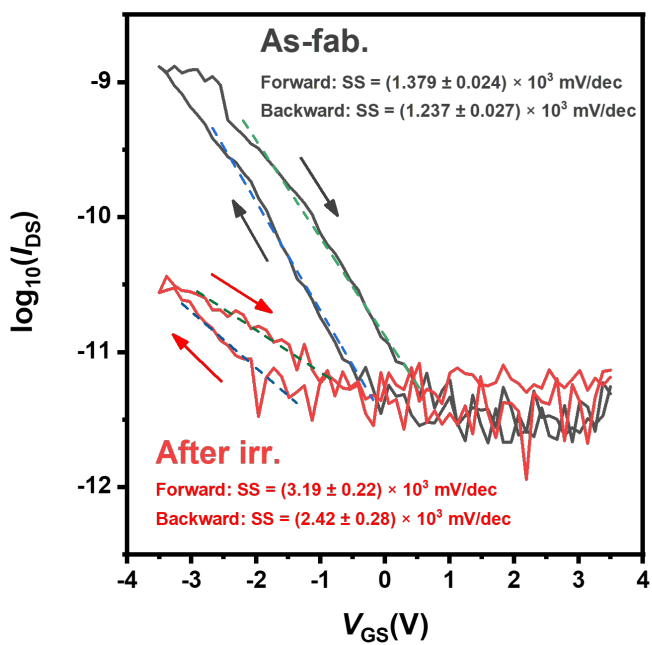

**Figure S11:** Log-scale transfer characteristics of the 9-AGNRFET measured before and after gamma irradiation. In the as-fabricated device, the extracted SS is $(1.379 \pm 0.024) \times 10^3$ mV/dec for the forward sweep and $(1.237 \pm 0.027) \times 10^3$ mV/dec for the backward sweep. After irradiation, the SS increases to $(3.19 \pm 0.22) \times 10^3$ mV/dec (forward sweep) and $(2.42 \pm 0.28) \times 10^3$ mV/dec (backward sweep). The dotted lines indicate the $V_{GS}$ ranges used for the linear fitting in the subthreshold regime. Arrows denote the gate-voltage sweep direction, allowing clear distinction between the forward and backward sweeps.

The SS was extracted from the linear region of the $\log_{10}(I_{DS})$–$V_{GS}$ characteristics.[9] Uncertainties in SS were obtained by propagating the standard errors of the linear-fit slopes through the SS definition.